\documentclass[12pt,usenames,dvipsnames,a4paper]{article}

\usepackage{soul}

\usepackage[utf8]{inputenc}
\usepackage[T1]{fontenc}

\usepackage{changepage}
\usepackage{amsmath,amsfonts,amssymb}
\usepackage{epsf,amsmath,bbold,amsfonts,stmaryrd}
\usepackage{mathrsfs}
\usepackage{appendix}
\usepackage{caption}
\usepackage{color}
\usepackage{datetime}
\usepackage{float}
\usepackage{graphicx}
\usepackage[colorlinks]{hyperref}
\hypersetup{pageanchor=false,citecolor=red,urlcolor=red}
\usepackage{indentfirst}
\usepackage[numbers,square,comma,sort&compress,merge]{natbib} 
\usepackage{subfig}
\numberwithin{equation}{section}
\usepackage{mathtools}
\usepackage{ytableau}
\usepackage{tabu}
\usepackage{esvect}

\usepackage{calc}

\allowdisplaybreaks

\def\a{\alpha}

\def\b{\beta}
\def\c{\chi}

\def\d{\delta}

\def\f{\frac}

\def\G{\Gamma}

\def\l{\left}

\def\m{\mu}
\def\n{\nu}

\def\p{\partial}

\def\r{\right}
\def\s{\sigma}

\def\x{\xi}

\def\z{\zeta}

\def\be{\begin{equation}}
\def\ee{\end{equation}}

\def\bea{\begin{eqnarray}}
\def\eea{\end{eqnarray}}

\def\ba{\begin{array}}
\def\ea{\end{array}}

\def\bc{\begin{center}}
\def\ec{\end{center}}

\def\bl{\begin{flushleft}}
\def\el{\end{flushleft}}

\def\br{\begin{flushright}}
\def\er{\end{flushright}}

\def\bi{\begin{itemize}}
\def\ei{\end{itemize}}

\def\bt{\begin{tabular}}
\def\et{\end{tabular}}

\makeatletter

\newsavebox\myboxA
\newsavebox\myboxB
\newlength\mylenA

\newcommand*\xoverline[2][0.75]{%
    \sbox{\myboxA}{$\m@th#2$}%
    \setbox\myboxB\null
    \ht\myboxB=\ht\myboxA`
    \dp\myboxB=\dp\myboxA%
    \wd\myboxB=#1\wd\myboxA
    \sbox\myboxB{$\m@th\overline{\copy\myboxB}$}
    \setlength\mylenA{\the\wd\myboxA}
    \addtolength\mylenA{-\the\wd\myboxB}%
    \ifdim\wd\myboxB<\wd\myboxA%
       \rlap{\hskip 0.5\mylenA\usebox\myboxB}{\usebox\myboxA}%
    \else
        \hskip -0.5\mylenA\rlap{\usebox\myboxA}
         {\hskip 0.5\mylenA\usebox\myboxB}%
    \fi}
\makeatother

\def\be{\begin{equation}}
\def\ee{\end{equation}}

\def\bea{\begin{eqnarray}}
\def\eea{\end{eqnarray}}

\def\f{\frac}

\def\p{\partial}

\usepackage{xspace}

\usepackage{xifthen}
\newcommand*\diff{\mathrm{d}} 
\newcommand*\ldiff[2][]{ \ifthenelse{\isempty{#1}}{ \diff
#2}{\diff^#1#2} \,} 
\let\limitint\int 
\renewcommand{\int}{\limitint \!} 

\begin{document}

\begin{titlepage}

\begin{adjustwidth}{-1.3cm}{-.7cm}
\begin{center}
    \bf \Large{The particle content of \texorpdfstring
     {$R^2$}~~gravity revisited}
\end{center}
\end{adjustwidth}

\begin{center}
\textsc{Georgios K. Karananas}
\end{center}

\begin{center}
\it {Arnold Sommerfeld Center\\
Ludwig-Maximilians-Universit\"at M\"unchen\\
Theresienstra{\ss}e 37, 80333 M\"unchen, Germany}
\end{center}

\begin{center}
\small
\texttt{\small georgios.karananas@physik.uni-muenchen.de} 
\end{center}

\begin{abstract}

Studying the spectrum of (pure) $R^2$ gravity on Minkowski background
inevitably results into a Catch-22: any consistent interpretation of its
particle dynamics dictates that no accidental gauge symmetries emerge, a
requirement that cannot be fulfilled when the theory is studied on
Minkowski.

For the case at hand, there is an emergent gauge redundancy corresponding to a
transverse-traceless shift of the graviton. This has detrimental consequences
since it empties the spectrum from all particle states. Being an artifact of
the linearized approximation on top of Minkowski background, the symmetry
does not persist at higher orders and degrees of freedom are reintroduced via
interactions, making $R^2$ gravity infinitely strongly-coupled.  

Provided that the theory is considered on appropriate backgrounds---for
instance de Sitter spacetime---or is supplemented with the Einstein-Hilbert
term, $R^2$ gravity propagates the two usual graviton polarizations plus one
additional (massless or massive) scalar field. We explicitly demonstrate that
in a fully covariant manner. 

\end{abstract}

\end{titlepage}

\section{Introduction}

In recent years there is renewed interest in alternatives to General
Relativity (GR), stemming mainly from cosmological considerations. Provided
that this is done in a selfconsistent manner, a zeroth-order approximation to
understanding a theory’s dynamics is to study its particle spectrum. The
textbook way to proceed is to work in the weak-field limit on some fixed
background. Usually, it suffices to consider excitations of the metric around
Minkowski-flat spacetime and retain the terms which are quadratic in the
perturbations so that the kinetic (and mass) term is identified. This works
remarkably well in GR and its higher-derivative generalizations~\cite
{Stelle:1977ry}, as well as in the Einstein-Cartan/Poincar\'e~\cite
{Neville:1978bk,Neville:1979rb,Sezgin:1979zf,Hayashi:1979wj,
Hayashi:1980qp,Karananas:2014pxa,Blagojevic:2018dpz,Lin:2018awc,Lin:2020phk,Barker:2024goa}
and metric-affine~\cite
{Percacci:2020ddy,Marzo:2021esg,Marzo:2021iok,Barker:2024ydb}
gravities, simply to name a few.

There is, however, an important point that should not be overlooked: in the
process of studying the spectrum, one needs to make sure that the linearized
action does not exhibit gauge invariance(s) other(s) than the one
(s) inherited from its parent non-linear theory---when talking about
conventional metrical gravity this is usually invariance under general
coordinate transformations (diffeomorphisms). 

The reason why one should worry about~\emph{accidental/emergent} gauge
invariances is rather obvious. Not being genuine redundancies, these only
manifest themselves under special circumstances---when a theory is linearized
on specific backgrounds and to leading order in the excitations~\cite
{Percacci:2020ddy}. Nevertheless they leave nontrivial imprints on the
dynamics by ``hiding'' (rather than eliminating) degrees of freedom. Therefore
it does not suffice to simply tread lightly when drawing conclusions about
the particle content of such systems. Any result is superficial and should
not be trusted.

In this paper we show that $R^2$-gravity in the metric formulation of
GR\,\footnote{For what matters, this holds true for all
(scalar-curvature)$^2$ gravities, irrespectively of the gravitational
formulation as we show in the follow-up paper~\cite{Karananas:tbp}.} falls
into this class of theories. Minkowski spacetime is a ``singular''
background, on top of which the quadratic action for the metric excitations
enjoys, apart from diffeomorphisms, a non-trivial accidental gauge invariance
that completely eliminates all degrees of freedom. Naturally, this symmetry
is absent at higher-order terms as well as when the theory is expanded on
different backgrounds such as de Sitter (dS).

This article is organized as follows. In Sec.~\ref{sec:EH}, we focus on the
Einstein-Hilbert (EH) action. We first rederive well-known results concerning
its flat spectrum. Next, we touch upon the accidental symmetries that the
theory exhibits at the leading, quadratic, order in the metric perturbations.
We show that these does not persist at higher orders, but this is not
problematic since they do not correspond to gauge redundancies. Sec.~\ref
{sec:R2}, is devoted to $R^2$-gravity. We demonstrate there in details that
linearizing on top of Minkowski, the quadratic action for the excitations
exhibits an accidental gauge invariance that forces all degrees of freedom to
be nondynamical. The fact that this symmetry is explicitly broken at higher
orders makes any statement concerning the particle spectrum based on such
considerations inconclusive. We then study the theory on top of dS spacetime
and (re-)confirm that it propagates the usual graviton and a massless scalar
field, as well known. This is due to the fact that the aforementioned gauge
redundancy is absent on ``healthy'' backgrounds. We also discuss what happens
when the action is supplemented with a linear in curvature term, in which
case Minkowski is well-defined. We conclude in Sec.~\ref
{sec:conclusions}. Throughout this article we work in a fully covariant
manner in order to systematically identify the gauge invariances, the
constraints that these impose on the energy-momentum tensor, and finally
elucidate the propagating degrees of freedom. When possible~(that is, for
Minkowski spacetime considerations) we utilize spin-projection operators.
Their definitions and properties can be found in Appendix~\ref
{app:spin_proj}.

\section{Einstein-Hilbert action}
\label{sec:EH}

\subsection{The particle spectrum}

Let us start our discussion by considering the Einstein-Hilbert action 
\be
\label{eq:EH_action_full} S^{\rm EH} = -\f{M_P^2}{2} \int \diff^4x \sqrt
 {g} R  \ ,
\ee 
where $M_P$ is the Planck mass, $g = -\det(g_{\m\n})$ is minus the
 determinant of the metric $g_{\m\n}$, and $R$ is the Ricci scalar that reads
\be
 R = g^{\m\n}\d^\rho_\lambda\left(\p_\rho \G^\lambda_{\m\n}-\p_\m\G^\lambda_
 {\rho\n} +\G^\lambda_{\rho\sigma}\G^\sigma_{\m\n}-\G^\lambda_
 {\m\sigma}\G^\sigma_{\rho\n} \right) \ ,
\ee 
with the Christoffel symbols given by
\be
\G^\lambda_{\m\n}= \f 1 2 g^{\kappa\lambda}\left(\p_\m g_
 {\kappa\n}+\p_\n g_{\kappa\m}-\p_\kappa g_{\m\n}\right) \ .
\ee

We expand the metric around Minkowski background
\be
\label{eq:metric_expansion} 
g_{\m\n} = \eta_{\m\n} + h_{\m\n}\ ,
\ee 
where $h_{\m\n}$ are perturbations and $\eta_{\m\n}={\rm diag}(1,-1,-1,-1)$ is
the flat metric. Plugging eq.~(\ref{eq:metric_expansion}) into~(\ref
{eq:EH_action_full}), and normalizing canonically the graviton 

\be
 \label{eq:graviton_canonical_norm}
h_{\m\n}~\mapsto~\f{2}{M_P} h_{\m\n} \ ,
 \ee
we end up with the usual Fierz-Pauli quadratic action for the excitations
\be
\label{eq:EH_action_lin} 
S^{\rm EH}_{2} = \f 1 2 \int \diff^4x \Big
 (\p_\rho h_{\m\n}\p^\rho h^{\m\n}-2\p^\m h_{\m\n}\p_\rho h^
 {\n\rho} +2\p^\m h \p^\n h_{\m\n} -\p_\m h \p^\m h\Big)\ .
\ee 
We adopted the standard shorthand notation $h=h^\m_\m$ for the trace of the
perturbation, and indexes are manipulated with the Minkowski metric.

To study the flat spectrum of the theory (and while doing so also identify its
gauge invariance(s)), it is most convenient and straightforward to break the
action down into decoupled spin subsectors; this is achieved by utilizing
projection operators~\cite
{barnes1965lagrangian,rivers1964lagrangian},~see also~\cite
{VanNieuwenhuizen:1973fi,Stelle:1977ry}. 

The metric perturbation $h_{\m\n}$, a symmetric rank-two tensor, is split into
a spin-2 transverse-traceless tensor (t), a spin-1 divergenceless vector
(v) and two spin-0 scalars (s,w)~\cite{VanNieuwenhuizen:1973fi}
\be 
\label{eq:metric_pert_decomp} 
h_{\m\n} = \left(P{_{\rm t}^2}_
 {\,\m\n\rho\sigma}+P{_{\rm v}^1}_{\,\m\n\rho\sigma}+ P{_{\rm s}^0}_
 {\,\m\n\rho\sigma}+P{_{\rm w}^0}_{\,\m\n\rho\sigma}\right)h^
 {\rho\sigma} \ .
\ee 
It is clear that $P{_{\rm a}^J}_{\,\m\n\rho\sigma}$  single out the spin-$J$
component(s) of $h_{\m\n}$. More details about the spin-projection operators
can be found in the Appendix~\ref{app:spin_proj}.~Before moving on let us
stress that the above~\emph{is not} the so-called St\"uckelberg trick, which
is not needed for the theories studied in this paper. Rather, 
Eq.~(\ref{eq:metric_pert_decomp}) is literally the tensorial analog of splitting
a vector into its transverse and longitudinal pieces---no degrees of freedom
nor gauge invariances others than the ones already present are introduced.

Plugging the decomposition~(\ref{eq:metric_pert_decomp}) into the linearized
action~(\ref{eq:EH_action_lin}) and using the properties of the
$P$-operators, we find that 
\begin{align} 
\label{eq:EH_action_projectors}
S_2^{\rm EH} = \int\diff^4x \Bigg[ &-\f 1 2h^{\m\n}\l(P{_
 {\rm t}^2}_{\,\m\n\rho\sigma}- 2 P{_{\rm s}^0}_
 {\,\m\n\rho\sigma} \r) \square h^{\rho\sigma} +\f{1}{M_P}P{_
 {\rm t}^2}_{\,\m\n\rho\sigma} h^{\m\n}T^{\rho\sigma} \nonumber \\
&+\f{1}{M_P} P{_{\rm s}^0}_{\,\m\n\rho\sigma} h^{\m\n}T^{\rho\sigma} +\f
   {1}{M_P}P{_{\rm v}^1}_{\,\m\n\rho\sigma} h^{\m\n}T^{\rho\sigma} +\f
   {1}{M_P} P{_{\rm w}^0}_{\,\m\n\rho\sigma} h^{\m\n}T^
   {\rho\sigma}\Bigg] \ ,
\end{align} 
where $\square=\p_\m\p^\m$ is the covariant d'Alembertian.

We have introduced an explicit coupling of the metric perturbation to a
(symmetric) energy-momentum tensor $T_{\m\n}$. The reason is that any
discussion concerning the spectrum requires that one studies the amplitude
for one-particle exchange, i.e. the propagator saturated with sources  
\be
\Pi = \mathcal O^{-1}_{\m\n\rho\sigma}T^{\m\n}T^{\rho\sigma} \ ,
\ee 
where $\mathcal O^{-1}$ is the inverse of the wave operator. 

Observe that only the tensorial and the s-scalar components of the metric
perturbation appear with kinetic terms in the action~(\ref
{eq:EH_action_projectors}). Owing to the orthogonality properties of the
projectors (remember, different spins do not mix at quadratic order), this
immediately translates into the following gauge invariance for the linearized
action
\be
\label{eq:gaug_inv_projectors}
\d h_{\m\n} = \l(P{_{\rm v}^1}_{\,\m\n\rho\sigma} + P{_{\rm w}^0}_
 {\,\m\n\rho\sigma} + P{_{\rm ws}^0}_{\,\m\n\rho\sigma}\r)\x^
 {\rho\sigma} \ ,
\ee 
provided that the
energy-momentum tensor satisfies
\be 
\label{eq:proj_constr}
P{_{\rm v}^1}_{\,\m\n\rho\sigma}T^{\rho\sigma} = 0 \ ,~~~P{_
 {\rm w}^0}_{\,\m\n\rho\sigma}T^{\rho\sigma} = 0 \ ,~~~P{_{\rm ws}^0}_
 {\,\rho\sigma\m\n}T^{\rho\sigma} =  0 \ . 
\ee 
Here, $\x_{\m\n}$ is a rank-two symmetric tensor, and $P{_{\rm ws}^0}$ a
projector (given in the Appendix~\ref{app:spin_proj}) connecting the s and
w spin-0 components. It is obvious that eq.~(\ref{eq:gaug_inv_projectors}) is
nothing more than linearized diffeomorphisms in the spin-projectors'
language. Indeed, the expression~(\ref{eq:gaug_inv_projectors}) can be
massaged into its standard form
\be
\label{eq:diffs}
\d h_{\m\n} = \p_\m \x_\n +\p_\n\x_\m \ ,
\ee 
with $\x^\m$ a four-vector. At the same time, the constraints~(\ref
{eq:proj_constr}) give rise to one and the same condition: the conservation
of $T_{\m\n}$
\be
\label{eq:e-m_tensor_cons}
\p_\m T^{\m\n} = 0 \ ,
\ee 
as it can be explicitly verified by using the definitions of the projectors. 

The lesson to be learned here is that the invariance under~(\ref
{eq:diffs}) and the associated source constraints~(\ref
{eq:e-m_tensor_cons}) arise from the non-appearance of the vectorial and
(one of the) scalar components of the graviton in the linearized EH action.
It is important to point out that the diffeomorphism invariance of~(\ref
{eq:EH_action_lin})  has been inherited from its parent theory (GR), see~
(\ref{eq:EH_action_full}). As such, the invariance under~(\ref{eq:diffs}), or
better say its nonlinear generalization 
\be
\label{eq:diff_full}
\d h_{\m\n} = \p_\m \x_\n + \p_\n \x_\m +\mathscr L_\x h_{\m\n} \ ,
\ee
with $\mathscr L_\x$ the Lie derivative, persists at all orders in the metric
perturbation around Minkowski spacetime.

Because of the universal coupling of the energy-momentum tensor to the metric
perturbation, all spin subsectors are affected and this has  important
consequences. Namely, being a gauge redundancy, it is responsible for
eliminating unphysical degrees of freedom from the spectrum of the theory. At
the end of the day, it is guaranteed that only the two polarizations
associated with the massless spin-2 graviton propagate (on-shell).

We now proceed with the inversion of the wave operator in~(\ref
{eq:EH_action_projectors}) to obtain the one-particle exchange amplitude. The
use of projectors makes this a trivial algebraic exercise; we easily find
that 
\be
\label{eq:EH_propagator_coordinate}
\Pi^{\rm EH} = -\f{P{_{\rm t}^2}_{\,\m\n\rho\sigma}- \f 1 2 P{_{\rm s}^0}_
 {\,\m\n\rho\sigma} }{\square} T^{\m\n} T^{\rho\sigma} \ .
\ee 
In momentum space, upon using the definitions of the $P$-operators from
Appendix~\ref{app:spin_proj} and imposing~(\ref{eq:e-m_tensor_cons}), we
obtain
\be
\Pi^{\rm EH}(k) = \f{T_{\m\n}(k)T^{\m\n}(-k) - \f 1 2 T (k)T (-k)}
 {k^2} \ ,
\ee 
with $T=T^\m_\m$ the trace of the energy-momentum tensor. This is the usual
expression for the massless graviton. Without loss of generality, we assume
propagation in the $z$-direction only so that the four-momentum is $k^\m =
(k^0,0,0,k^3)$. On-shell $k^2=0$ and $T_{0\n}(k) = -T_{3\nu}(k)$ as follows
from the momentum-space counterpart of~(\ref{eq:e-m_tensor_cons}), so the
above yields
\be
\Pi^{\rm EH}(k) = \lim_{k^2\to 0} \f{\f 1 2\l(T_{11}(k)-T_{22}(k)\r)\l(T_{11}
 (-k)-T_{22}(-k)\r) + 2T_{12}(k)T_{12}(-k)}{k^2} \ .
\ee 
from which we observe that the propagator contains only the helicity $\pm 2$
graviton polarizations, as it should. 

\subsection{Accidental symmetries of the quadratic action and their fate
 beyond leading order}

It so happens that linearized gravity at the quadratic level enjoys~\emph
{accidental invariances}. Capturing the dynamics of a free massless
spin-2 field, the action~(\ref{eq:EH_action_lin}) is invariant under the shift
of the graviton
\be
\label{eq:dc_metric}
\d_c h_{\m\n}= c_{\m\n} \ ,
\ee
by a constant $c_{\m\n}$. 

Moreover, since the action does not contain parameters carrying dimension of
mass, it is trivially invariant under global scale transformations
(dilatations)\,\footnote{Unlike free theories of lower-spin fields, it is
well-known~\cite{Anselmi:1999bu} (see also~\cite{Farnsworth:2021zgj}) that
the Fierz-Pauli action fails to be invariant under the full conformal group
if the graviton transforms as a spin-2 primary. }
\be
\label{eq:dlambda_metric}
\d_\lambda h_{\m\n} = -\l(x^\rho\p_\rho+1\r)h_{\m\n} \ ,
\ee
as it can be readily verified.

Let us now argue that the accidental symmetries~(\ref{eq:dc_metric},\ref
{eq:dlambda_metric}) associated with the free character of the action~(\ref
{eq:EH_action_lin}) cannot persist beyond the leading order in excitations.
To this end, we expand the gravitational action~(\ref{eq:EH_action_full}) to
cubic order in $h_{\m\n}$. In terms of the canonically normalized graviton
(\ref{eq:graviton_canonical_norm}) we find 
\begin{eqnarray} 
S_3^{\rm EH} &\propto&\f{1}{M_P} \int \diff^4 x \Bigg
 ( \f{1}{4}h^{\rho\s}\p_\rho h^{\m\n}\p_\s h_
 {\m\n}- \f{1}{4}h^{\rho\s}\p_\rho h\p_\s h + h^
 {\n\rho}\p_\rho h \p_\m h_\n^\m-\f{1}{2} h^
 {\m\n} \p_\rho h\p^\rho h_{\m\n} \nonumber \\
&& + \f{1}{8}h \p_\m h \p^\m h - h^
      {\m\n}\p_\rho h^\rho_\m \p_\s h_\n^\s-h^
      {\m\n}\p_\n h_\m^\rho \p_\s h_\rho^\s +\f{1}
      {2} h \p_\m h^{\m\n}\p_\rho h_\n^\rho \nonumber \\
&& + \f 1 2h^{\m\n} \p^\rho h_{\m\n}\p_\s h_\rho^\s - \f
      {1}{4}h\p_\m h \p_\n h^{\m\n}+\f 1 2 h^
      {\m\n} \p_\rho h_{\n\s}\p^\s h_\m^\rho + \f 1 2 h^
      {\m\n} \p_\rho h_
      {\n\s}\p^\rho h_\m^\s \nonumber \\
&& -\f{1}{4}h\p_\rho h_{\m\n} \p^\n h^{\m\rho} - \f{1}
     {8}h\p_\rho h^{\m\n} \p^\rho h_{\m\n}\Bigg) \ .
\end{eqnarray}

Observe that both the shift and dilatational symmetries are explicitly broken:
the former due to the fact that the graviton's self-interaction terms are of
the form $h\p h \p h$, and the latter because all higher-dimensional pieces
stemming from the expansion of the EH action appear together with inverse
powers of the dimensionful Planck mass $M_P$.

It should be noted that being global invariances, they are not capable of
eliminating degrees of freedom from the theory. Thus, their breakdown is not
problematic since no new degrees of freedom, pathological or not, are
(re-)introduced.

\section{\texorpdfstring{$R^2$}~~gravity}
\label{sec:R2}

\subsection{Particle spectrum and accidental gauge invariance(s)}

Having fully set the stage with the EH action, let us move to $R^2$ gravity
\be 
\label{eq:metric_action_full} S^{\rm R^2} = - \int \diff^4x \sqrt
 {g} R^2 \ .
\ee
Expanding the metric around Minkowski background, see Eq.~(\ref
{eq:metric_expansion}), we end up with the following quadratic action for the
perturbations (see also~\cite{Alvarez-Gaume:2015rwa,Hell:2023mph})
\be
\label{eq:metric_action_lin} S^{\rm R^2}_{2} = - \int \diff^4 x\left
 (\p_\m\p_\n h^{\m\n}-\square h \right)^2 \ .
\ee

A simple inspection of the above raises immediately a red flag. There is no
way it captures the dynamics of a spin-2 field---\emph{at best}, it may
describe a spin-0 system. The reason is that in addition to the
diffeomorphisms~(\ref{eq:diffs}), the theory~\emph{around Minkowski spacetime
enjoys an accidental gauge redundancy}. Indeed, $S^{\rm R^2}_2$ is invariant
under the following gauge transformation
\be
\label{eq:acc_1} 
\d_{\rm t}h_{\m\n} = \z^{\rm TT}_{\m\n} \ ,
\ee 
with $\z^{\rm TT}_{\m\n}$ a transverse-traceless tensor
\be
\p^\m \z^{\rm TT}_{\m\n}=\eta^{\m\n}\z^{\rm TT}_{\m\n} = 0 \ .
\ee

The crux of it all is that this emergent symmetry is an artifact of
linearizing on top of this specific background and this background only. It
does not survive when considering higher-orders in $h_{\m\n}$ nor different
backgrounds. Let us also stress that since~(\ref{eq:acc_1}) is gauged it is
fundamentally different from the accidental symmetries enjoyed by the EH
action, and as such it has a twofold effect on the system. First, it
completely eliminates from the quadratic action the spin-2 excitations and
second, it constraints non-trivially the energy-momentum tensor. The latter,
we shall see soon, actually nullifies completely all particle dynamics on top
of Minkowski.

As before, we proceed by decomposing the action into independent spin
subsectors in terms of the projection operators introduced earlier. Plugging
the resolution~(\ref{eq:metric_pert_decomp}) of $h_{\m\n}$ into the
linearized action~(\ref{eq:metric_action_lin}) supplemented with a source
term for the field, we find that only the scalar s-component of the metric
perturbation has a kinetic term 
\begin{align} 
S^{\rm R^2}_2 = -\int \diff^4 x \Bigg[ 3 P{_{\rm s}^0}_
 {\,\m\n\rho\sigma}&h^{\m\n}\square^2 h^{\rho\sigma}+P{_{\rm s}^0}_
 {\,\m\n\rho\sigma} h^{\m\n}T^{\rho\sigma} \nonumber \\
&+ P{_{\rm t}^2}_{\,\m\n\rho\sigma} h^{\m\n}T^{\rho\sigma} +P{_
   {\rm v}^1}_{\,\m\n\rho\sigma} h^{\m\n}T^{\rho\sigma} + P{_
   {\rm w}^0}_{\,\m\n\rho\sigma} h^{\m\n}T^{\rho\sigma}\Bigg] \ ,
\end{align} and moreover it is of fourth order in derivatives. At first sight
 this appears to be problematic, for it signals the presence of a scalar
 ghost. However, appearances are deceiving: the above describes a theory that
 contains no propagating degrees of freedom, as correctly pointed out
 in~\cite{Hell:2023mph}. Let us elaborate on this. In addition to the usual
 conservation~(\ref{eq:e-m_tensor_cons}) of the energy-momentum tensor owing
 to the invariance of the action under the (linearized) diffeomorphisms~(\ref
 {eq:diffs}), the absence of a kinetic term for the spin-2 component of $h_
 {\m\n}$ implies the presence of an extra gauge invariance
\be
\d_{\rm t} h_{\m\n} = P{_{\rm t}^2}_{\,\m\n\rho\sigma}\x^{\rho\sigma} \ ,  
\ee which is exactly the one we found previously~(\ref{eq:acc_1}) by
 inspecting the action---now, it effortlessly emerges in the spin-projection
 formalism. In turn, this translates into an extra condition for $T_
 {\m\n}$. Specifically, it dictates that the energy-momentum tensor, in
 addition to~(\ref{eq:proj_constr}), be also subject to
\be
\label{eq:em_transverse}
P{_{\rm t}^2}_{\,\m\n\rho\sigma} T^{\rho\sigma} = 0 \ .
\ee 
In momentum space and for propagation in the $z$-direction, this means that
the energy-momentum tensor is forced to assume the following form\,\footnote
{Notice that the trace of the energy-momentum tensor is proportional to
$k^2$, so it vanishes on shell. }
\be
\label{eq:em_explicit_components}
T_{\m\n}= T_{00}\begin{pmatrix}
1 & 0& 0& -\f{k^0}{k^3}\\
0 & \f{k^2}{(k^3)^2}&0&0\\
0 & 0 &\f{k^2}{(k^3)^2}&0 \\
-\f{k^0}{k^3} & 0 & 0 & \f{(k^0)^2}{(k^3)^2}
\end{pmatrix} \ ,~~~k^2=(k^0)^2-(k^3)^2 \ .
\ee
As a result, the saturated propagator 
\be
\Pi^{\rm R^2}(k)\propto \f {P{_{\rm s}^0}_
 {\,\m\n\rho\sigma}}{k^4}T^{\m\n}(k)T^{\rho\sigma}(-k)\propto \f{T(k)T(-k)}{k^4} \propto \f{T_{00}(k)T_{00}(-k)}{(k^3)^4}    \ ,
\ee 
has no pole, signaling that the scalar is not propagating  and thus the flat
particle spectrum of the theory is trivial. The result that, in spite of
appearances, $R^2$ gravity does not propagate any degrees of freedom was
obtained for the first time recently~\cite{Hell:2023mph} in a non-covariant
manner.

Before moving on, let us point out that what we encounter here bears a
superficial resemblance to the massless 3-form theory (see for instance~\cite
{Dvali:2005an}), or more precisely its dual counterpart. Similar analysis
shows that the scalar mode in the aforementioned theory also presents itself
in the action with a quartic derivative kinetic term~\cite
{Hell:2023mph,Golovnev:2023zen}. Nevertheless, the propagator of the field
also does not have a pole so the spectrum of the 3-form theory does not
contain dynamical degrees of freedom; this has been re-confirmed in~\cite
{Barker:2024juc} with the use of the PSALTer software. Contrary to $R^2$
gravity, however, the gauge redundancy responsible for trivializing the
spectrum is a genuine feature of the full theory and not an overlooked
artifact of the linearization on top of some inadmissible background.

\subsection{Higher order terms and the breakdown of the accidental gauge
 invariance}

As we stressed before, the fact that on top of Minkowski background and to
leading order in the metric perturbation $R^2$-gravity does not propagate
particles is due to the accidental tensorial gauge invariance~(\ref
{eq:acc_1})---emerging in the linearized approximation---and its associated
stringent condition~(\ref{eq:em_transverse}) on the energy-momentum tensor.
Consequently, although the statement is algebraically correct, conceptually
it is misguided and therefore no trustworthy conclusion concerning the
particle spectrum should be drawn by this result. Studying the spectrum of
$R^2$ gravity on top of Minkowski spacetime is, simply put,
inconsistent.\footnote{An analogous strong-coupling problem resulting from
curvature-squared terms in the context of Einstein-Cartan gravity was pointed
out in~\cite{Barker:2023fem}.}

Perhaps not surprisingly, at higher orders in $h_{\m\n}$ the action is not
invariant under the gauged transverse-traceless shift~(\ref{eq:acc_1}) of the
graviton. This is an immediate aftermath of the fact that the parent
nonlinear theory is only invariant under general coordinate transformations.
This can be seen by plugging the decomposition~(\ref
{eq:metric_expansion}) into the action~(\ref{eq:metric_action_full}) and
expanding to cubic order in $h_{\m\n}$, yielding\,\footnote{Keep in mind that
$h_{\m\n}$ in $R^2$ gravity is dimensionless.}
\be 
 S = S_2 + S_3   \ ,
\ee 
where the quadratic piece $S_2$ is given in eq.~(\ref
{eq:metric_action_lin}), while 
\begin{align}
& S_3 = - \int\diff^4x  (\p_\m \p_\n h^{\m\n}-\square h) \Big(2h_{\rho\sigma}
  (\p^\rho\p^\sigma h-2 \p^\rho\p_\lambda h^{\sigma\lambda}+\square h^
  {\rho\sigma}) + \f 1 2 h(\p^\rho\p^\sigma h_
  {\rho\sigma}-\square h) \nonumber \\
&\qquad\qquad~\,+2 \p^\rho h\left(\p^\sigma h_{\rho\sigma}-\f 1 4 \p_\rho
  h\right)-2 \p^\rho h_{\rho\sigma}\p_\lambda h^
  {\sigma\lambda} -\p^\rho h^{\sigma\lambda}\p_\lambda h_
  {\rho\sigma} +\f 3 2 \p_\lambda h_{\rho\sigma} \p^\lambda h^
  {\rho\sigma}\Big) \ .
\end{align}

The response of $S_3$ to the tensorial gauge symmetry~(\ref{eq:acc_1}), after
some algebra is found to not vanish nor combine into a full derivative
\begin{align}
\d_{\rm t} S_3 \propto \int \diff^4x (\p_\m \p_\n h^{\m\n}-\square h) \Big(2\z^
 {\rm TT}_{\rho\sigma}(\p^\rho\p^\sigma h&-2 \p^\rho\p_\lambda h^
 {\sigma\lambda}+\square h^{\rho\sigma}) +2h^{\rho\sigma}\square \z^
 {TT}_{\rho\sigma} \nonumber \\
&\qquad-2 \p_\rho \z^{\rm TT}_{\sigma\lambda}\p^\lambda h^
  {\rho\sigma} + 3 \p_\lambda\z^{\rm TT}_{\rho\sigma}\p^\lambda h^
  {\rho\sigma} \Big) \ ,
\end{align} 
signaling the breakdown of the accidental symmetry of the quadratic
action.\footnote{It is important to stress that one~\emph{cannot use the
equations of motion}~when checking the invariance of the action.} 

As a result, the constraint~(\ref{eq:em_transverse}) on the energy-momentum
tensor is completely meaningless, rendering the analysis of the spectrum
inconclusive. The bottom line is that any consistent study of the particle
dynamics of the theory requires that in one way or another the accidental
gauge symmetry be eliminated from the quadratic action. For the pure $R^2$
gravity~(\ref{eq:metric_action_full}) this means that Minkowski spacetime
must be excluded altogether as a background.

\subsection{The way out: \texorpdfstring{$R^2$}~~gravity on non-flat
 backgrounds}

The selfconsistent way to conclude about the particle content of pure $R^2$
gravity is provided by linearizing the theory on top of some background in
which the gauge invariance~(\ref{eq:acc_1}) is absent, e.g. dS spacetime. It
will become clear that the particle spectrum of $R^2$-gravity is not linearly
empty. 

Let us expand the metric around a dS spacetime
\be
\label{eq:metr_dS} 
g_{\m\n} = \bar g_{\m\n} + h_{\m\n} \ ,
\ee 
with $\bar g_{\m\n}$ the background metric and as before $h_
{\m\n}$ are small perturbations. The quadratic part of~(\ref
{eq:metric_action_full}) evaluated on~(\ref{eq:metr_dS}) reads
\begin{eqnarray}
&&\bar S_2 = \int \diff^4 x \sqrt{\bar g} \l [\f{\bar R}{2}\l
  (\mathcal D_\rho h_{\m\n}\mathcal D^\rho h^{\m\n}-2\mathcal D^\m h_
  {\m\n}\mathcal D_\rho h^{\n\rho} +\mathcal D^\m h \mathcal D^\n h_
  {\m\n} \r) \r.\nonumber\\
&&\qquad\qquad\qquad\qquad\qquad\l. +\f{\bar R^2}{12}\l( h_{\m\n}h^
   {\m\n}-\f {h^2}{4} \r) -\l(\mathcal D_\m\mathcal D_\n h^
   {\m\n}-\mathcal D^2 h\r)^2 \r]  , 
\end{eqnarray} 
where $\bar g$ is (minus) the determinant of the dS metric, $\mathcal D_\m$
the covariant derivative, $\mathcal D^2=\bar g^{\m\n}\mathcal D_\m\mathcal
D_\n$ the dS d'Alembertian, $h=\bar g^{\m\n}h_{\m\n}$, and $\bar R=12\Lambda$
the curvature of the background. Now, indexes are manipulated with $\bar g_
{\m\n}$ and its inverse. 

Although the form of the above is already suggestive, we can go a step further
and recast the action as~\cite{Alvarez-Gaume:2015rwa} 
\be 
\label{eq:act_dS_linearized}
\bar S_2 = \bar S^{\rm EH}_2 +\bar S^{\rm R^2}_2 \ ,
\ee
where 
\begin{align}
\label{eq:EH_dS_linearized}
\bar S^{\rm EH}_2 = \int \diff^4 x \sqrt{\bar g} \Bigg [\f{\bar R}{2}\big(\mathcal D_\rho h_{\m\n}\mathcal D^\rho h^{\m\n}-2\mathcal D^\m h_
  {\m\n}\mathcal D_\rho h^{\n\rho} +2&\mathcal D^\m h \mathcal D^\n h_
  {\m\n}-\mathcal D_\m h \mathcal D^\m
  h\big) \nonumber \\
  &~~+\f{\bar R^2}{12}\l( h_{\m\n}h^{\m\n}+\f{h^2}{2} \r) \Bigg] \ ,
\end{align}
corresponds to the Einstein-Hilbert action linearized on top of dS,
\footnote
{Indeed, take the Einstein-Hilbert action given previously in~(\ref
{eq:EH_action_full}) and supplement it with a cosmological constant $\Lambda$ 
\be 
\label{eq:action_EH_CC}
S^{\rm EH}_{\rm \Lambda}=-\f{M_P^2}{2}\int\diff^4x \sqrt{g}\l( R -6 \Lambda \r) \ .
\ee 
Using
\be
g_{\m\n} = \bar g_{\m\n} + \f{4\sqrt{3 \Lambda}}{M_P} h_{\m\n} \ ,~~~\Lambda = \f{\bar R}{12}  \ ,
\ee
and expanding~(\ref{eq:action_EH_CC}) to second order in $h_{\m\n}$ yields~(\ref
{eq:EH_dS_linearized}).} 
and 
\be
\bar S^{\rm R^2}_2 = -  \int \diff^4 x \sqrt{\bar g} \l(\mathcal D_\m\mathcal D_\n h^
   {\m\n}-\mathcal D^2 h-\f{\bar R}{4} h\r)^2 \ ,
\ee
which is the dS-analog of~(\ref{eq:metric_action_lin}). 

Let us now discuss the gauge redundancies of~(\ref
{eq:act_dS_linearized}). First, it is obvious that since the background is an
Einstein space, under dS diffeomorphisms 
\be
\d h_{\m\n} = \mathcal D_\m \x_\n +\mathcal D_\n \x_\m \ ,
\ee 
the action is invariant
\be
\d \bar S_2 = 0 \ . 
\ee

Next, it is also obvious that in the presence of $\bar S^{\rm EH}_2$,  the
tensorial accidental symmetry~(\ref{eq:acc_1}) is absent
\be
\d_{\rm t} \bar S_2 \propto  \int\diff^4x\sqrt{\bar g} \,\bar R\, h^{\m\n}\l(\mathcal D^2 -\f{\bar R}
 {6}\r)\z_{\m\n}^{\rm TT} \neq 0 \ ,
\ee 
and so the theory certainly propagates the usual transverse-traceless piece of
the metric perturbation corresponding to the massless spin-2 graviton. 

As we shall now show, $R^2$-gravity also propagates an additional massless
(for dimensional reasons) spin-0 field whose dynamics is associated with the
higher derivative part of the action. Since we cannot use spin-projectors
(at least not in a straightforward manner) in dS, we employ the standard
auxiliary-field method to isolate the scalar mode. To this end we  introduce
a Lagrange multiplier, say $\c$, and rewrite the linearized action~(\ref
{eq:act_dS_linearized}) as
\begin{eqnarray}
\label{eq:auxiliary_linear}
&&\bar S_2= \int \diff^4 x \sqrt{\bar g} \l [\f{\bar R}{2}\l(\mathcal D_\rho
  h_{\m\n}\mathcal D^\rho h^{\m\n}-2\mathcal D^\m h_
  {\m\n}\mathcal D_\rho h^{\n\rho} +2\mathcal D^\m h \mathcal D^\n h_
  {\m\n} -\mathcal D_\m h \mathcal D^\m
  h\r) \qquad\qquad\r.\nonumber\\
&&~\qquad\qquad\qquad\qquad\l. +\f{\bar R^2}{12}\l( h_{\m\n}h^
   {\m\n}+\f{h^2}{2} \r) -\c\l(\mathcal D_\m\mathcal D_\n h^
   {\m\n}-\mathcal D^2 h-\f{\bar R}{4} h\r)+\f{\c^2}{4} \r]  .
\end{eqnarray} 
One can verify that on the equations of motion of $\c$ the above boils down to
what we started with. 

Then, to get rid of the kinetic mixing between $\c$ and $h_{\m\n}$  we perform
the following change of variables (which is nothing more than a linearized
Weyl transformation of the metric together with a trivial rescaling of the
graviton)
\be 
\c = \sqrt{\f{2\bar R}{3}}\hat \c \ ,~~~h_{\m\n} =\f{1}{\sqrt{\bar R}}\l( \hat h_{\m\n} - \f{1}{\sqrt 6}\bar g_
 {\m\n}\hat \c\r) \ ,
\ee 
to obtain 
\begin{eqnarray}
&&\bar S_2 = \f 1 2  \int \diff^4 x \sqrt{\bar g} \Bigg [\mathcal D_\rho \hat h_{\m\n}\mathcal D^\rho \hat h^
  {\m\n}-2\mathcal D^\m \hat h_{\m\n}\mathcal D_\rho \hat h^
  {\n\rho} +2\mathcal D^\m \hat h \mathcal D^\n \hat h_
  {\m\n} -\mathcal D_\m \hat h \mathcal D^\m \hat
  h \nonumber\\
&&\qquad\qquad\qquad\qquad\qquad\qquad\qquad\qquad\qquad +\f{\bar R}{6}\l( \hat h_
   {\m\n}\hat h^{\m\n}+\f{\hat h^2}{2} \r) + \p_\m \hat \c \p^\m \hat \c\Bigg] \ ,
\end{eqnarray}
which demonstrates that the theory propagates the two spin-2 polarizations of
the graviton plus a massless scalar field.

Our analysis exhibits that $R^2$ gravity is (at least classically) completely
equivalent to standard GR with a~\emph{nonvanishing}~cosmological constant
$\Lambda = \bar R/12$ plus a massless scalar field; this statement can be
easily shown to persist in the full non-linear theory, see e.g.~\cite
{Alvarez-Gaume:2015rwa}. 

\subsection{Yet another way out: \texorpdfstring{$R+R^2$}~~on flat
 background}

If one insists on not giving up on Minkowski spacetime as a background, the
simplest way (at least we could think of) is to radically deviate from $R^2$
gravity by supplementing the action~(\ref{eq:metric_action_full}) with the
Einstein-Hilbert term~(\ref{eq:EH_action_full}). The latter effectively plays
the same role as the  curvature $\bar R$ of dS: it explicitly breaks the
unwanted emergent symmetry responsible for trivializing the spectrum. 

Consider  
\be
\label{eq:Starobinsky_action}
S' = S^{\rm EH} - \f{1}{12 f^2} S^{\rm R^2} \ ,
\ee
with $f$ a (real) dimensionless constant. Upon using~(\ref
{eq:EH_action_lin}) and expressing~(\ref{eq:metric_action_lin}) in terms of
the canonically-normalized graviton~(\ref{eq:graviton_canonical_norm}), we
obtain  
\begin{align}
\label{eq:metric_RR2_full} 
S'_2 =  \f 1 2 \int \diff^4x \Bigg[\p_\rho h_{\m\n}\p^\rho h^
  {\m\n}-2\p^\m h_{\m\n}\p_\rho h^{\n\rho} &+2\p^\m h \p^\n h_
  {\m\n} -\p_\m h \p^\m h  \nonumber\\
  &+ \f{2}{3f^2M_P^2}\left(\p_\m\p_\n h^{\m\n}-\square h \r)^2 \Bigg] \ .
\end{align} 
Having studied the spectrum on top of dS, and because of the striking
similarity of the above to~(\ref{eq:act_dS_linearized}), we already have a
strong hint about the particle content when it comes to $R+R^2$ gravity: it
propagates a massless spin-2 graviton and a~\emph{massive}~---as we shall
shortly show---spin-0 field. In terms of the spin-projection operators the
above action assumes the form
\begin{align} 
S'_2 = -\f{1}{2} &\int \diff^4x  \Bigg[h^{\m\n}\l(P{_
 {\rm t}^2}_{\,\m\n\rho\sigma}- 2 P{_{\rm s}^0}_{\,\m\n\rho\sigma}\l
 (1+\f{\square}{f^2M_P^2}\r) \r) \square h^{\rho\sigma} \nonumber \\
&+P{_{\rm t}^2}_{\,\m\n\rho\sigma} h^{\m\n}T^{\rho\sigma} + P{_
  {\rm s}^0}_{\,\m\n\rho\sigma} h^{\m\n}T^{\rho\sigma} +P{_{\rm v}^1}_
  {\,\m\n\rho\sigma} h^{\m\n}T^{\rho\sigma} + P{_{\rm w}^0}_
  {\,\m\n\rho\sigma} h^{\m\n}T^{\rho\sigma}\Bigg] \ ,
\end{align} 
which shows the intricate dynamics of the scalar sector of the theory. This is
even more clearly reflected in the saturated propagator 
\be
\label{eq:prop_RR2}
\Pi' = \Pi^{\rm EH} + \Pi^{\rm s} \ ,
\ee
that comprises two terms: $\Pi^{\rm EH}$ that is given in~(\ref
{eq:EH_propagator_coordinate}) and corresponds to the massless spin-2
graviton, while
\be
\label{eq:propRR22}
\Pi^{\rm s} \propto - \f{P{_{\rm s}^0}_{\,\m\n\rho\sigma}}{\square + f^2 M_P^2} T^{\m\n} T^{\rho\sigma} \ ,
\ee
corresponds to a healthy spin-0 field with mass $m_s=fM_P$.~This is
Starobinsky's scalaron~\cite{Starobinsky:1980te}, which is well known to be
present in~(\ref{eq:Starobinsky_action}). 

Equivalently, and in an attempt to be maximally pedagogic here, instead of
utilizing projectors we could have used the same trick as in dS with an
auxiliary field $\c$ to rewrite~(\ref{eq:metric_RR2_full}) as
\begin{align}
S'_2 = \f 1 2 \int \diff^4x \Bigg[\p_\rho h_{\m\n}\p^\rho h^
  {\m\n}-2\p^\m h_{\m\n}&\p_\rho h^{\n\rho} +2\p^\m h \p^\n h_
  {\m\n} -\p_\m h \p^\m h \nonumber \\
  &+ \sqrt{\f{2}{3}}\f{\c}{f M_P}\left(\p_\m\p_\n h^{\m\n}-\square h \r) -\f{\c^2}{4} \Bigg] \ .
\end{align}
The change of variables 
\be
\c = 2fM_P \hat \c \ ,~~~ h_{\m\n} =\hat h_{\m\n} + \f{1}{\sqrt 6} \eta_{\m\n} \hat \c \ ,
\ee
diagonalizes the kinetic sector and the action becomes
\begin{align}
S'_2 = \f 1 2 \int \diff^4x \Bigg[\p_\rho \hat h_{\m\n}\p^\rho \hat h^
  {\m\n}-2\p^\m \hat h_{\m\n}\p_\rho \hat h^{\n\rho} +2\p^\m \hat h \p^\n \hat h_
  {\m\n} &-\p_\m \hat h \p^\m \hat h \nonumber \\
  &+ \p_\m \hat \c \p^\m \hat \c - f^2 M_P^2\hat \c^2 \Bigg] \ ,
\end{align} 
in full agreement with what we expect from the propagator~(\ref
{eq:prop_RR2},\ref{eq:propRR22}). Just like the dS considerations, there is
no difficulty in generalizing the above result at the nonlinear level~\cite
{Kehagias:2013mya}.

\section{Conclusion}
\label{sec:conclusions}

The purpose of this paper was modest. In a nutshell, our main findings can be
summarized as follows. 

For $R^2$ gravity, Minkowski spacetime is an ill-defined background, in the
sense that on top of it the linearized quadratic action exhibits a nontrivial
accidental gauge redundancy (other than diffeomorphism invariance). This is
responsible for emptying the spectrum from all particle states. Importantly,
the emergent symmetry is bound to be explicitly broken at some order in
perturbation theory. This can be trivially ``proven'' by contradiction: if
this is not the case, then the full nonlinear action must exhibit such a
symmetry, which is of course not true. Therefore, no conclusion about the
degrees of freedom propagated by $R^2$ gravity based on the Minkowski
computation should be drawn. Let us mention that this was also realized
in~\cite{Golovnev:2023zen}, from a different viewpoint. 

The selfconsistent way to proceed is by considering backgrounds that break
such unwanted emergent gauge symmetries. For pure $R^2$ gravity this is for
example de Sitter spacetime. We (re-)confirmed that the spectrum of the
theory is of course not empty, but rather comprises the graviton and a
massless scalar field. 

\section*{Acknowledgements} 

I am grateful to Will Barker, Gia Dvali, Alex Kehagias, Misha Shaposhnikov and
Sebastian Zell for valuable discussions, and to Sebastian Zell for important
comments on the manuscript.

\appendices

\section{Spin-projection operators}
\label{app:spin_proj}

The building blocks of the spin-projection operators are the usual
transverse and longitudinal projectors
\be
\Theta_{\m\n} = \eta_{\m\n} -\f{\p_\m\p_\n}{\square} \ ,~~~\Omega_
 {\m\n} = \f{\p_\m\p_\n}{\square} \ ,
\ee 
that are complete 
\be
\Theta_{\m\n} + \Omega_{\m\n} = \eta_{\m\n} \ ,
\ee
and orthonormal
\be
\Theta_{\m\a}\Theta^\a_{\n} = \Theta_{\m\n} \ ,~~~\Omega_
 {\m\a}\Omega^\a_\n=\Omega_{\m\n}\ ,~~~\Theta_{\m\a}\Omega^\a_\n =
 0 \ .
\ee 

Taking appropriate tensor products of $\Theta_{\m\n}$ and $\Omega_
{\m\n}$, operators that project out the spin-$J$ components of the metric
perturbation are constructed. The ones used in the main text are given
by~\cite{rivers1964lagrangian,barnes1965lagrangian,VanNieuwenhuizen:1973fi} 
\begin{eqnarray}
&&P{_{\rm t}^2}_{\,\m\n\rho\sigma} = \f 1 2 \left(\Theta_{\m\rho}\Theta_
  {\n\sigma}+\Theta_{\m\sigma}\Theta_{\n\rho}\right) - \f 1 3 \Theta_
  {\m\n}\Theta_{\rho\sigma} \ ,\label{eq:spin2_proj}\\
&&P{_{\rm v}^1}_{\,\m\n\rho\sigma} = \f 1 2 \left(\Theta_{\m\rho}\Omega_
  {\n\sigma}+\Theta_{\m\sigma}\Omega_{\n\rho}+\Theta_{\n\rho}\Omega_
  {\m\sigma}+\Theta_{\n\sigma}\Omega_{\m\rho}\right) \ ,\\
&&P{_{\rm s}^0}_{\,\m\n\rho\sigma} = \f 1 3 \Theta_{\m\n}\Theta_
  {\rho\sigma} \ ,\\
&&P{_{\rm w}^0}_{\,\m\n\rho\sigma} = \Omega_{\m\n} \Omega_
  {\rho\sigma} \ \label{eq:spin0_proj}. 
\end{eqnarray} 
The above sum up to unity in the space of symmetric rank-two tensors  
\be 
P^2_{\m\n\rho\sigma}+P^1_{\m\n\rho\sigma}+ P{_{\rm s}^0}_
 {\,\m\n\rho\sigma}+P{_{\rm w}^0}_{\,\m\n\rho\sigma} = \f 1 2\left
 ( \eta_{\m\rho}\eta_{\n\sigma}+\eta_{\m\sigma}\eta_
 {\n\rho} \right) \ ,
\ee 
as they should.

Owing to the presence of the two scalar (s,w) components, it is necessary to
also introduce two ``off-diagonal'' projectors that implement mappings
between the spin-$0$ subspace. These are 
\be 
\label{eq:offdiagonal_proj}
P{_{\rm sw}^0}_{\,\m\n\rho\sigma} = \sqrt{\f 1 3}\Theta_
 {\m\n}\Omega_{\rho\sigma} \ ,~~~P{_{\rm ws}^0}_
 {\,\m\n\rho\sigma} = \sqrt{\f 1 3}\Omega_{\m\n}\Theta_
 {\rho\sigma} \ . 
\ee 
Finally, the full set of projectors~(\ref{eq:spin2_proj})-(\ref
{eq:spin0_proj}) and~(\ref{eq:offdiagonal_proj}) are orthonormal in the
following sense 
\be 
P{_{ac}^{J'}}_{\,\m\n\alpha\beta} P{_{db}^{J}}^{~\a\b}_
 {~~~\rho\sigma}= \d_{cd}\d^{J'J}P{_{ab}^{J}}_
 {~\m\n\rho\sigma} \ .
\ee

\bibliographystyle{utphys}
\bibliography{Refs}

\end{document}